\documentclass[twocolumn,aps,showpacs,floatfix,prc]{revtex4}
\usepackage[dvips]{epsfig}
\usepackage{subeqn}

\begin{document}

\title{ Nuclear reaction rates and the primordial nucleosynthesis }

\author{ Abhishek Mishra$^1$ and D. N. Basu$^2$ }

\affiliation{Variable Energy Cyclotron Centre, 1/AF Bidhan Nagar, Kolkata 700 064, India }

\email[E-mail 1: ]{abhishek.mishra@veccal.ernet.in}
\email[E-mail 2: ]{dnb@veccal.ernet.in}

\date{\today }

\begin{abstract}

    The theoretical predictions of the primordial abundances of elements in the big-bang nucleosynthesis (BBN) are dominated by uncertainties in the input nuclear reaction rates. We investigate the effect of modifying these reaction rates on light element abundance yields in BBN by replacing the thirty-five reaction rates out of the existing eighty-eight. We have studied these yields as functions of evolution time or temperature. We find that using these new reaction rates results in only a little increase in helium mass fraction over that obtained previously in BBN calculations. This allows insights into the role of the nuclear reaction rates in the setting of the neutron-to-proton ratio during the BBN epoch. We observe that even with considerable nuclear physics uncertainties, most of these nuclear reactions have minimal effect on the standard BBN abundance yields of $^6$Li and $^7$Li. 

\vspace {0.25cm}
\end{abstract}

\maketitle

\noindent
\section{Introduction}
\label{section1}

    Primordial nucleosynthesis took place just a few moments after the big-bang \cite{Ho64} and the universe evolved very rapidly allowing only the synthesis of the lightest nuclides such as D, $^{3,4}He$ and $^{6,7}Li$. In addition to these stable nuclei some unstable, or radioactive, isotopes like tritium or ${^3}H$ and $^{7,8}Be$ were also produced during the big-bang nucleosynthesis (BBN). These unstable isotopes either decayed or fused with other nuclei to make one of the stable isotopes. It lasted for only about seventeen minutes (during the period from three to about twenty minutes from the beginning of space expansion) and after that, the temperature and density of the universe fell below that which is required for nuclear fusion and prevented elements heavier than beryllium from forming while at the same time allowing unburned light elements, such as deuterium, to exist. The abundances of these nuclides are probes of the conditions of the universe during the very early stages of its evolution. The conditions during the BBN are believed to be well described in terms of standard models of cosmology, nuclear and particle physics, which determine the values of temperature, nucleon density, expansion rate, neutrino content, neutrino-antineutrino asymmetry, etc. Deviations from the BBN test the parameters of these models and constrain nonstandard physics or cosmology that may alter the conditions during BBN \cite{Wa67,Wa69,Io09}. Sensitivity to several parameters and physics input in the BBN model have been investigated thoroughly in the past \cite{No00,Cy02,Cy04,Cy08,Se04,Fu09}. 

    The nuclear reaction rates $<\sigma v>$ in reaction network calculations, where $\sigma$ is the nuclear-fusion cross section and $v$ is the relative velocity between the participant nuclides, are the most important inputs for modeling the BBN and stellar evolution. Although $v$ is well described by a Maxwell- Boltzmann velocity distribution for a given temperature $T$, the cross section $\sigma$ can only be obtained from laboratory experiments, some of which are not as well known as desired \cite{No00,Cy02,Cy04,Cy08,Se04}. Several factors influence the measured values of the cross sections and the theoretical estimates of the thermonuclear reaction rates depend on the various approximations used. In the network calculations one needs to account for the Maxwellian-averaged thermonuclear reaction rates and difference \cite{Fo88,An99} in these rates affects the description of elemental synthesis in the BBN or in stellar evolution. In this work, we consider the effect the nuclear reaction rates on the primordial abundances of elements. 

\noindent
\section{Big-bang nucleosynthesis reaction network}
\label{section2}

    The predictions of the standard BBN theory depend on low energy nuclear cross sections and on three additional parameters, the number of flavours of light neutrinos ($N_\nu$), the neutron lifetime ($\tau_n$) and the ratio of baryons to photons ($\eta=n_B/n_\gamma$) in the universe \cite{Co95,Sh95}. In its standard $N_\nu=3.0$ form, primordial nucleosynthesis is, therefore, a one parameter theory, depending only on $\eta$ as $\tau_n=885.7(8) s$ is an accurately measured quantity. One of the most essential predictions of the standard big-bang model is the synthesis of light elements in the primordial universe. There are three pillars of big-bang cosmology which are the Hubble expansion, cosmic background radiation (CBR) and the BBN. The BBN examines back to earlier times of the universe than the other two and it also has to deal with nuclear and particle physics along with cosmology. Although Hubble expansion can also be used by some other alternative cosmological theories (e.g. steady state), the evidences of CBR and BBN observations lead cosmologists to a universe which was very hot and dense in the very beginning. The Friedmann-Robertson-Walker cosmological model is the standard scenario underlying the BBN theory. Moreover, the solution of the Einstein equations lead to an isotropic and homogeneous universe, so the uniformity of the CBR temperature, which is $T= 2.7277 \pm 0.002^o$K across the sky, as well as the success of the standard BBN theory serve to validate this approximation. The primordial harvest of light elements is determined by the time during the expansion of the universe. It is possible to characterize the BBN in general, that the paradigm most frequently uses the Friedmann equation to relate the big-bang expansion rate, H, to the thermal properties of the particles present at that epoch. During expansion, the rates of the weak interactions that transform neutrons and protons, and the rates of the nuclear reactions that build up the complex nuclei are involved. 

    The reactions which happened at the duration of BBN can be organised into two groups, firstly, the reactions that interconvert neutrons and protons which are $n+e^+ \leftrightarrow p+\bar\nu_e$, $p+e^- \leftrightarrow n+\nu_e$ and $n \leftrightarrow p+e^-+\bar\nu_e$ and secondly, the rest of the reactions. The first group can be expressed in terms of the mean neutron lifetime and the second group is determined by many different nuclear cross section measurements. The formation of deuterium begins in the process of $p+n \leftrightarrow D + \gamma$. This reaction is exothermic with an energy difference of 2.2246 MeV, but since photons are 10$^9$ times more numerous than protons, the reaction does not proceed until the temperature of the expanding universe falls to about 0.3 MeV, when the photo-destruction rate is lower than the production rate of deuterons. When the deuteron formation starts some further reactions proceed to make ${^4}He$ nuclei. 

\begin{eqnarray}
 D + n \rightarrow {^3}H + \gamma \nonumber \\
 {^3}H + p \rightarrow {^4}He + \gamma \nonumber \\
 D + p \rightarrow {^3}He + \gamma \nonumber \\
 {^3}He + n \rightarrow {^4}He + \gamma
\label{seqn1}
\end{eqnarray}
\noindent
Both light helium ${^3}He$ and normal helium ${^4}He$ are formed along with the ${^3}H$. Since helium nucleus binding energy is 28.3 MeV and more bound than the deuterons and the temperature has already fallen to 0.1 MeV, these reactions can be photoreactions and only go one way. The following four reactions also produce ${^3}He$ and ${^4}He$ and they usually go faster since they do not involve the relatively slow process of photon emission. 

\begin{eqnarray}
 D + D \rightarrow {^3}He + n \nonumber \\
 D + D \rightarrow {^3}H + p \nonumber \\
 {^3}He + D \rightarrow {^4}He + p \nonumber \\
 {^3}H + D \rightarrow {^4}He + n
\label{seqn2}
\end{eqnarray}
\noindent
Eventually the temperature gets so low that the electrostatic repulsion of the deuterons and other charged particles causes the reactions to stop. The deuteron:proton ratio when the reactions stop is quite small, and essentially inversely proportional to the total density of protons and neutrons (to be precise, goes like the -1.6 power of the density). Almost all the neutrons in the universe end up in normal helium nuclei. For a neutron:proton ratio of 1:7 at the time of deuteron formation, 25$\%$ of the mass ends up in helium. Deuterium peaks around 100 seconds after the big-bang, and is then rapidly swept up into helium nuclei. A very few helium nuclei combine into heavier nuclei giving a small abundance of ${^7}Li$ coming from the big-bang. ${^3}H$ decays into ${^3}He$ with a 12 year half-life so no ${^3}H$ survives to the present, and ${^7}Be$ decays into ${^7}Li$ with a 53 day half-life and also does not survive. The uncertainties for the reactions ${^3}He+{^4}He \rightarrow {^7}Be+\gamma$, ${^3}H+ {^4}He \rightarrow {^7}Li+\gamma$ and $p+{^7}Li \rightarrow {^4}He+{^4}He$ may lead to about 50$\%$ uncertainty in the predicted yield of ${^7}Li$. In the present work, we replace thirty-five Maxwellian-averaged thermonuclear reaction rates by new ones in the Kawano/Wagoner BBN code \cite{Ka92} and study its effect on the primordial abundances of elements. 

\noindent
\section{ Thermonuclear reaction rates }
\label{section3}

    The twelve most important nuclear reactions which affect the predictions of the abundances of the light elements [${^4}He$, D, ${^3}He$, ${^7}Li$]  are $n-$decay, $p(n,\gamma)d$, $d(p,\gamma){^3}He$, $d(d,n){^3}He$, $d(d,p)t$, ${^3}He(n,p)t$, $t(d,n){^4}He$, ${^3}He(d,p){^4}He$, ${^3}He(\alpha,\gamma){^7}Be$, $t(\alpha,\gamma){^7}Li$, ${^7}Be(n,p){^7}Li$ and ${^7}Li(p,\alpha){^4}He$. Instead of cross sections $\sigma$, the nuclear reaction inputs to BBN take the form of thermal rates. These rates are computed by averaging nuclear reaction cross sections over a Maxwell-Boltzmann distribution of energies. The Maxwellian-averaged thermonuclear reaction rate $<\sigma v>$ at some temperature $T$, is given by the following integral \cite{Bo08}:

\begin{equation}
 <\sigma v> = \Big[\frac{8}{\pi\mu (k_B T)^3 } \Big]^{1/2} \int \sigma(E) E \exp(-E/k_B T) dE,
\label{seqn3}
\end{equation}
\noindent
where $E$ is the centre-of-mass energy, $v$ is the relative velocity and $\mu$ is the reduced mass of the reactants. At low energies (far below Coulomb barrier) where the classical turning point is much larger than the nuclear radius, barrier penetrability can be approximated by $\exp(-2\pi\zeta)$ so that the charge induced cross section can be decomposed into

\begin{equation}
 \sigma(E) = \frac{S(E)\exp(-2\pi\zeta)}{E}
\label{seqn4}
\end{equation}
\noindent
where $S(E)$ is the astrophysical $S$-factor and $\zeta$ is the Sommerfeld parameter, defined by

\begin{equation}
 \zeta = \frac{Z_1Z_2e^2}{\hbar v}
\label{seqn5}
\end{equation}
\noindent
where $Z_1$ and $Z_2$ are the charges of the reacting nuclei in units of elementary charge $e$. Except for narrow resonances, the $S$-factor $S(E)$ is a smooth function of energy, which is convenient for extrapolating measured cross sections down to astrophysical energies. In the case of a narrow resonance, the resonant cross section $\sigma_r(E)$ is generally approximated by a Breit-Wigner expression. 

\begin{table}[htbp]
\vspace{0.0cm}
\centering
\caption{\label{tab:table1} The nuclear reactions with modified thermonuclear reaction rates for BBN are tabulated along with the applicable $T_9$ (in units of $10^9$$^o$K) ranges and references. Rows without $T_9$ ranges show applicability for the entire range.}
\begin{tabular}{ccc}
\hline
\hline
$d(p,\gamma){^3}He$ & $T_9 \le 0.8$ \cite{An04} & $T_9 > 0.8$ \cite{An99} \\
$d(d,n){^3}He$ & $T_9 \le 3.0$ \cite{An04} & $T_9 > 3.0$ \cite{An99} \\
$d(d,p)t$ & $T_9 \le 3.0$ \cite{An04} & $T_9 > 3.0$ \cite{An99} \\
$d(\alpha,\gamma){^6}Li$ & \cite{An99}\\
$t(d,n){^4}He$ & $T_9 \le 0.5$ \cite{An04} & $T_9 > 0.5$ \cite{An99} \\
$t(\alpha,\gamma){^7}Li$ & $T_9 \le 8.0$ \cite{An04} & $T_9 > 8.0$ \cite{An99} \\
${^3}He(n,p)t$ & $T_9 \le 3.0$ \cite{An04} & $T_9 > 3.0$ \cite{Sm93} \\
${^3}He(d,p){^4}He$ & $T_9 \le 2.0$ \cite{An04} & $T_9 > 2.0$ \cite{Sm93} \\
${^3}He({^3}He,2p){^4}He$ & \cite{An99}\\
${^3}He(\alpha,\gamma){^7}Be$ & $T_9 \le 8.0$ \cite{An04} & $T_9 > 8.0$ \cite{An99} \\
${^4}He(\alpha n,\gamma){^9}Be$ & \cite{An99}\\
${^4}He(\alpha \alpha,\gamma){^{12}}C$ & \cite{An99}\\
${^6}Li(p,\gamma){^7}Be$ & \cite{An99}\\
${^6}Li(p,\alpha){^3}He$ & \cite{An99}\\
${^7}Li(p,\alpha){^4}He$ & $T_9 \le 7.0$ \cite{An04} & $T_9 > 7.0$ \cite{An99} \\
${^7}Li(\alpha,\gamma){^{11}}B$ & \cite{An99}\\
${^7}Be(n,p){^7}Li$ & $T_9 \le 0.2$ \cite{An04} & $T_9 > 0.2$ \cite{Sm93} \\
${^7}Be(p,\gamma){^8}B$ & \cite{An99}\\
${^7}Be(\alpha,\gamma){^{11}}C$ & \cite{An99}\\
${^9}Be(p,\gamma){^{10}}B$ & \cite{An99}\\
${^9}Be(p,d \alpha){^4}He$ & \cite{An99}\\
${^9}Be(p,\alpha){^6}Li$ & \cite{An99}\\
${^9}Be(\alpha,n){^{12}}C$ & \cite{An99}\\
${^{10}}B(p,\gamma){^{11}}C$ & \cite{An99}\\
${^{10}}B(p,\alpha){^7}Be$ & \cite{An99}\\
${^{11}}B(p,\gamma){^{12}}C$ & \cite{An99}\\
${^{11}}B(p,\alpha\alpha){^4}He$ & \cite{An99}\\
${^{12}}C(p,\gamma){^{13}}N$ & \cite{An99}\\
${^{12}}C(\alpha,\gamma){^{16}}O$ & \cite{An99}\\
${^{13}}C(p,\gamma){^{14}}N$ & \cite{An99}\\
${^{13}}C(\alpha,n){^{16}}O$ & \cite{An99}\\
${^{13}}N(p,\gamma){^{14}}O$ & \cite{An99}\\
${^{14}}N(p,\gamma){^{15}}O$ & \cite{An99}\\
${^{15}}N(p,\gamma){^{16}}O$ & \cite{An99}\\
${^{15}}N(p,\alpha){^{12}}C$ & \cite{An99}\\
\hline
\hline
\end{tabular} 
\vspace{0.0cm}
\end{table}

    The neutron induced reaction cross sections at low energies can be written as \cite{Bl55} 

\begin{equation}
 \sigma(E) = \frac{R(E)}{v}
\label{seqn6}
\end{equation}
\noindent
facilitating extrapolation of the measured cross sections down to astrophysical energies, where $R(E)$ is a slowly varying function of energy \cite{Mu10} and is similar to an $S$-factor.

    The compilations concerning specifically standard BBN reaction rates were performed by  Caughlan et al. \cite{Fo88} and Smith et al. \cite{Sm93}. We are using rates for some BBN reactions from the newest compilations by Angulo et al. \cite{An99} and Descouvemont et al. \cite{An04} which are meant to supersede the earlier compilations. For two-body reactions the rates are computed from Eq.(3). The $S$-factor is a smooth function of energy except for narrow resonances. When it is assumed to be constant, the integrand in Eq.(3) is peaked at a most effective energy which can be approximated by a Gaussian function and the integral in Eq.(3) can be calculated analytically. In \cite{An99}, this approximation is not used and the integral is solved numerically for non-resonant as well as broad resonant contributions. Also a detailed analysis of uncertainties is provided for more realistic lower and upper bounds to the adopted rates. The analysis in \cite{Sm93} was performed using polynomial expansions for the cross-sections, and the uncertainties on rates were in general only estimated by allowing $S$-factor limits to encompass all existing data whereas in the work of \cite{An04} the cross-sections are analyzed in the R-matrix framework which provides a more rigorous energy dependence, based on Coulomb functions. In addition to that the advantage of all the available data is taken to constrain the $S$-factor, not restricting the data sets only to the energy range of BBN. Also careful evaluation of the uncertainties associated with the cross-sections and reaction rates is done based on standard statistical techniques. More accurate results are, therefore, expected. The Maxwellian-averaged thermonuclear reaction rates of relevance in astrophysical plasmas are evaluated \cite{An99} assuming either that the target nucleus is in its ground state, or that the target states are thermally populated following a Maxwell-Boltzmann distribution, except some cases involving isomeric states. Table-I lists nuclear reactions for which Maxwellian-averaged thermonuclear reaction rates are modified \cite{An99,An04} for BBN. 
 
\begin{figure}[htbp]
\eject\centerline{\epsfig{file=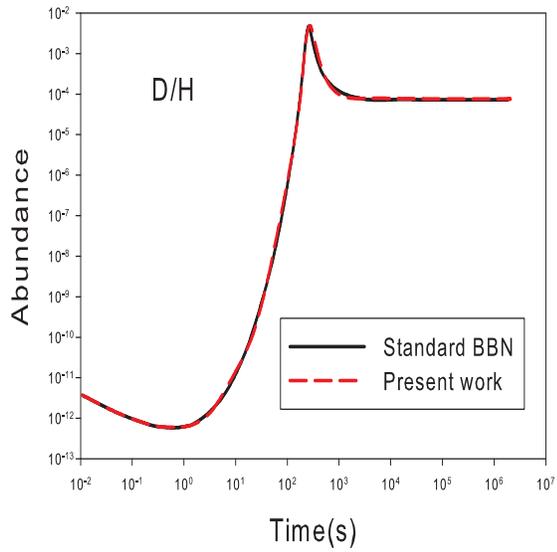,height=7.25cm,width=7.25cm}}
\caption
{(Color online) Plots of the deuterium abundances versus time. The continuous line represents the standard BBN results whereas the dashed line represents the same with modified Maxwellian-averaged thermonuclear reaction rates \cite{An99}.}
\label{fig1}
\vspace {5.2cm}
\end{figure}

\begin{figure}[htbp]
\eject\centerline{\epsfig{file=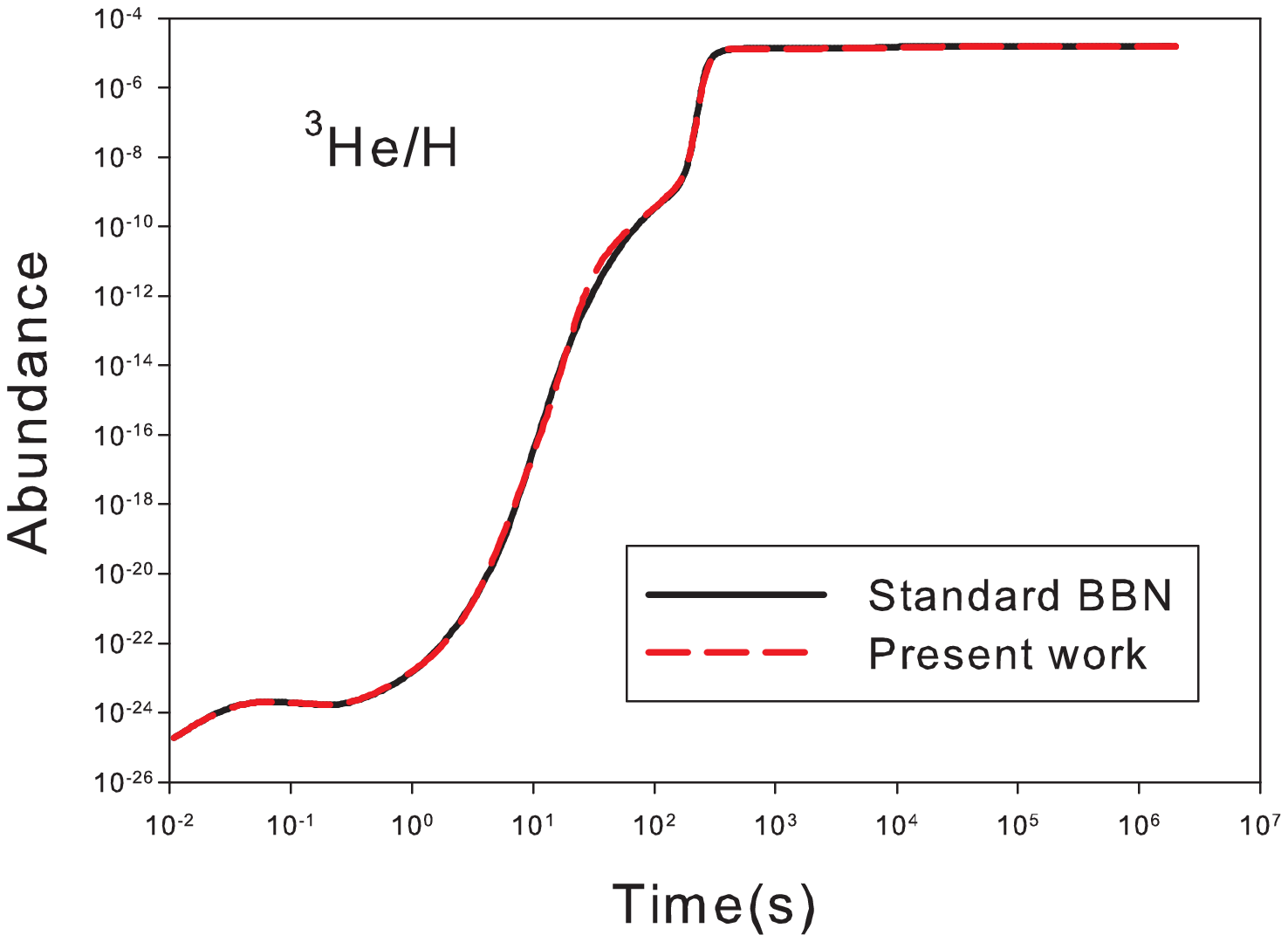,height=7.25cm,width=7.25cm}}
\caption
{(Color online) Same as Fig.1 but for the ${^3}He$ abundances. }
\label{fig2}
\vspace {0.0cm}
\end{figure}

\begin{figure}[htbp]
\eject\centerline{\epsfig{file=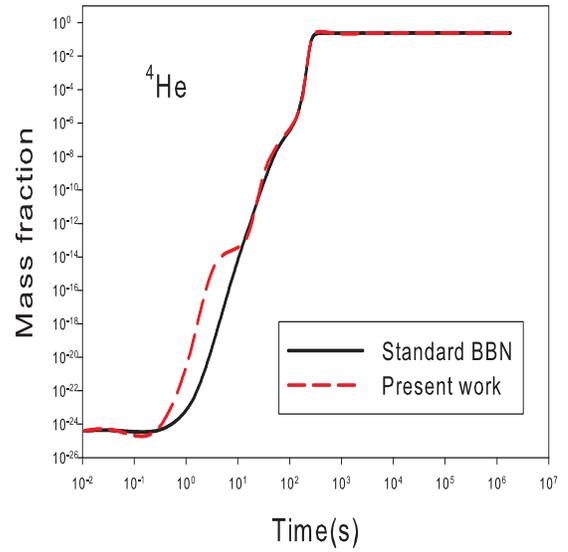,height=7.25cm,width=7.25cm}}
\caption
{(Color online) Same as Fig.1 but for ${^4}He$ mass fractions.}
\label{fig3}
\vspace {5.4cm}
\end{figure}

\begin{figure}[htbp]
\eject\centerline{\epsfig{file=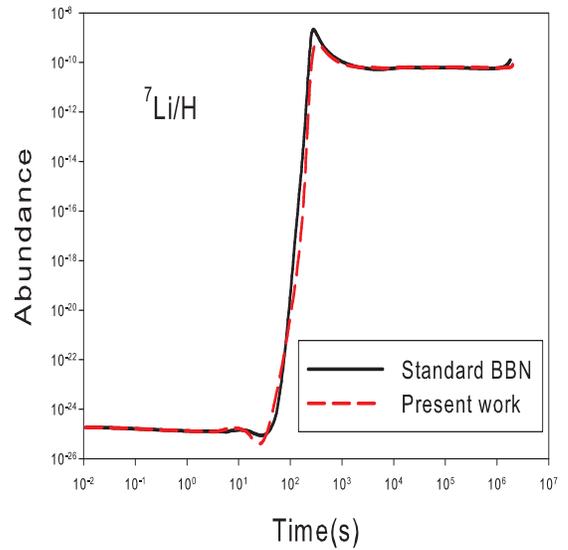,height=7.25cm,width=7.25cm}}
\caption
{(Color online) Same as Fig.1 but for the ${^7}Li$ abundances.}
\label{fig4}
\vspace {0.0cm}
\end{figure}

\begin{figure}[htbp]
\eject\centerline{\epsfig{file=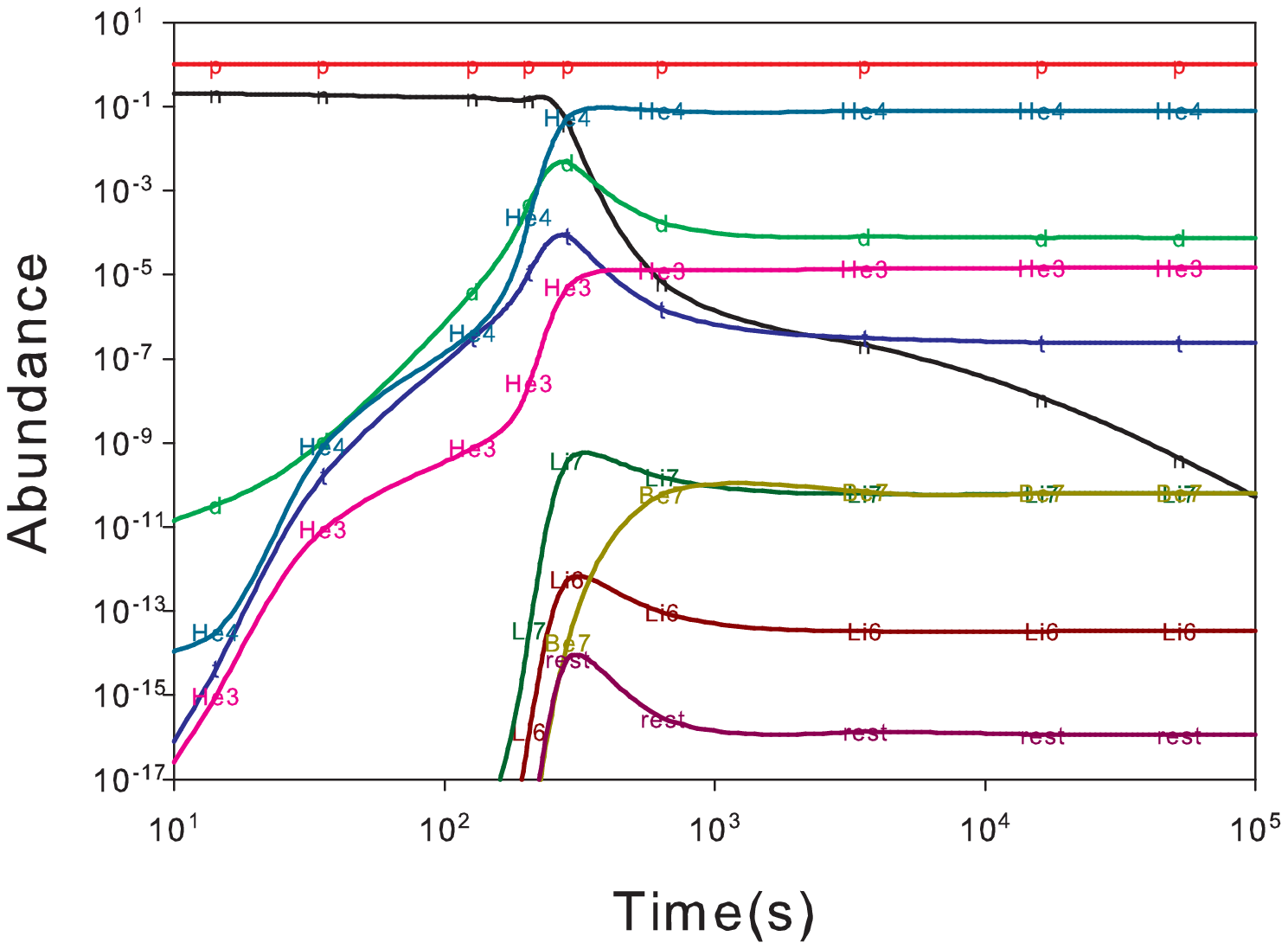,height=8cm,width=8cm}}
\caption
{(Color online) Plots of all the abundances with modified Maxwellian-averaged thermonuclear reaction rates \cite{An99}. The curve marked `rest' represents sum of all the abundances of nuclei higher than ${^7}Be$.}
\label{fig5}
\vspace {0.0cm}
\end{figure}     
    
\noindent
\section{ Effect of thermonuclear reaction rates on primordial abundances }
\label{section4}
 
    A comprehensive study of the effect of nuclear reaction rates on primordial nucleosynthesis is performed. In Fig.1 and Fig.2 the primordial abundances $D/H$ and ${^3}He/H$ of $D$ and ${^3}He$ nuclei with respect to the number of $H$ nuclei are shown as functions of evolution time. The continuous line represents the standard BBN results whereas the dashed line represents the same with modified Maxwellian-averaged thermonuclear reaction rates \cite{An99}. In Fig.3 the ${^4}He$ primordial mass fractions are plotted against the evolution time for standard BBN and with the modified reaction rates \cite{An99}. In Fig.4 the primordial abundances ${^7}Li/H$ of ${^7}Li$ nuclei with respect to the number of $H$ nuclei are shown as functions of evolution time for standard BBN and with the modified reaction rates \cite{An99}. The plots of calculations incorporating R-matrix reaction rates \cite{An04} are not shown in the figures as they are limited to only small $T_9$ values and do not alter the results. In Fig.5 plots of abundances of all the elements with respect to the number of $H$ nuclei are shown as functions of evolution time. The impact on BBN of the recent compilation of thermonuclear reactions rates \cite{An99} does not make large overall changes. These modified rates do affect the magnitude of these predictions at intermediate times for cases such as ${^4}He$ or ${^7}Li$. However, as obvious from the figures the final values of these predictions remain almost same.
         
\begin{table}[htbp]
\vspace{0.0cm}
\centering
\caption{\label{tab:table1} Yields at the test ($\eta_{10} = 3.162$) and the WMAP ($\eta_{10} = 6.19 \pm 0.15$) \cite{WMAP} baryonic densities.}
\begin{tabular}{ccccccc}
\hline
\hline
 &Kawano-&This work&Kawano-&This work&Observations&Factor \\ 
 &   Wagoner  &                &Wagoner&&&\\ \hline
 &   [Test]&[Test]&[WMAP]&[WMAP]&& \\ \hline

${^4}He$&0.2410&0.2411&0.2479&0.2479&0.232-0.258\cite{Ol04}&$\times 10^{0}$ \\
$D/H$&7.250&7.277&2.519&2.563&$2.82^{+0.20}_{-0.19}$\cite{Pe08}&$\times 10^{-5}$ \\
${^3}He/H$&1.546&1.613&1.033&1.058&0.9-1.3\cite{Ba02}&$\times 10^{-5}$ \\
${^7}Li/H$&1.268&1.367&4.627&5.019&1.1$\pm$0.1\cite{Ho09}&$\times 10^{-10}$ \\ \hline
\hline
\end{tabular} 
\vspace{0.0cm}
\end{table}     
    
    All these calculations described so far for the standard BBN and with the modified reaction rates \cite{An99} were performed with a test value for the ratio of the baryons to photons $\eta = \eta_{10} \times 10^{-10} = 3.162 \times 10^{-10}$ which reproduces the observed ${^7}Li$ abundance for the standard BBN. In Table-II, the comparison between BBN abundances deduced using the test value ($\eta_{10}=3.162$) and from the WMAP \cite{WMAP} results ($\eta_{10}=6.19$) is provided. In this comparison, the results corresponding to the present work use all the thirty-five modified reaction rates \cite{An99,An04} listed in Table-I. We find that using these new reaction rates results in insignificant increase in helium mass fraction over that obtained previously in standard BBN calculations. One observes that even with considerable nuclear physics uncertainties, most of these nuclear reactions have minimal effect on the standard BBN abundance yields of ${^7}Li$. 

\section{ Summary and conclusion }
\label{section5}

    In summary, we find little effect on the standard BBN abundance yields by replacing the Maxwellian-averaged thermonuclear reaction rates by new ones, even given a fair uncertainty in issues that bear on key reaction rates. The addition of some new reactions to the BBN code had virtually no effect on the BBN abundances \cite{Bo10}. The chances of solving either of the ''lithium problems'' by conventional nuclear physics means are unlikely and, if these problems stand up to future observations, we may be forced into just such non-standard BBN scenarios. At present, however, theoretical predictions of the primordial $^6$Li abundance are extremely uncertain due to difficulties in both theoretical estimates and experimental determinations of the $^2$H($\alpha,\gamma$)$^6$Li radiative capture reaction cross section. We also argue that present observational capabilities do not yet allow the detection of primeval $^6$Li in very metal-poor stars of the galactic halo. However, if the critical cross section is very high in its plausible range and the baryon density is relatively low, then improvements in $^6$Li detection capabilities may allow the establishment of $^6$Li as another product of BBN. It is also noted that a primordial $^6$Li detection could help resolve current concerns about the extragalactic $D/H$ determination. It could be conjectured, however, that the new thermonuclear reaction rates used here may turn out to be important for nonstandard BBN scenarios \cite{Do93,Do94} with new particle physics.


\newpage
\noindent
\begin{thebibliography}{99}

\bibitem{Ho64} F. Hoyle and R. J. Tayler, Nature (London) {\bf 203}, 1108 (1964).

\bibitem{Wa67} R. Wagoner, W. A. Fowler, and F. Hoyle, Astrophys. J. {\bf 148}, 3 (1967).

\bibitem{Wa69} R. Wagoner, Astrophys. J. Supp. {\bf 18}, 247 (1969).

\bibitem{Io09} F. Iocco, G. Mangano, G. Miele, O. Pisanti, and P. D. Serpico, Phys. Rep. {\bf 472}, 1 (2009).

\bibitem{No00} K. M. Nollett and S. Burles, Phys. Rev. {\bf D 61}, 123505 (2000).

\bibitem{Cy02} R. H. Cyburt, B. D. Fields, and K. A. Olive, Astropart. Phys. {\bf 17}, 87 (2002).

\bibitem{Cy04} R. H. Cyburt, Phys. Rev. {\bf D 70}, 023505 (2004).
 
\bibitem{Cy08} R. H. Cyburt, B. D. Fields and K. Olive, J. Cosm. Astropart. Phys. {\bf 11}, 12 (2008).

\bibitem{Se04} P. D. Serpico, S. Esposito, F. Iocco, G. Mangano, G. Miele, and O. Pisanti, J. Cosmol. Astropart. Phys. {\bf 12}, 010 (2004).

\bibitem{Fu09} G. M. Fuller and C. J. Smith, arXiv:1009.0277.

\bibitem{Fo88} G. R. Caughlan, and W. A. Fowler, Atom. Data Nucl. Data Tables {\bf 40}, 283 (1988).

\bibitem{An99} C. Angulo et al., Nucl. Phys. {\bf A 656}, 3 (1999).

\bibitem{Co95} C. J. Copi, D. N. Schramm and M. S. Turner, Science {\bf 267}, 192 (1995).

\bibitem{Sh95} C. J. Copi, D. N. Schramm and M. S. Turner, Phys. Rev. Lett. {\bf 75}, 3981 (1995).

\bibitem{Ka92} L. Kawano, FERMILAB Report No. PUB-92/04-A, January 1992 (unpublished).

\bibitem{Bo08} R. N. Boyd, {\it An Introduction to Nuclear Astrophysics} (University of Chicago, Chicago, 2008), 1st ed.

\bibitem{An04} P. Descouvemont, A. Adahchour, C. Angulo, A. Coc, E. Vangioni-Flam, Atom. Data Nucl. Data Tables {\bf 88}, 203 (2004).

\bibitem{Sm93} M.S. Smith, L.H. Kawano and R.A. Malaney, Astrophys. J. Suppl. {\bf 85}, 219 (1993).

\bibitem{Bl55} J. M. Blatt and V. F. Weisskopf, {\it Theoretical Nuclear Physics} (John Wiley $\&$ Sons, New York; Chapman $\&$ Hall Limited, London.) 

\bibitem{Mu10} Tapan Mukhopadhyay, Joydev Lahiri and D. N. Basu, Phys. Rev. {\bf C 82}, 044613 (2010); {\it ibid} Phys. Rev. {\bf C 83}, 039902(E) (2011).

\bibitem{WMAP} E. Komatsu et al. [WMAP Collaboration], Astrophys. J. Suppl. {\bf 192}, 18 (2011)
[arXiv:1001.4538].

\bibitem{Ol04} K. A. Olive and E. Skillman, Astrophys. J. {\bf 617}, 290 (2004).

\bibitem{Pe08} M. Pettini {\it et al.}, Mon. Not. R. Astron. Soc. {\bf 391}, 1499 (2008).

\bibitem{Ba02} T. Bania, R. Rood and D. Balser, Nature {\bf 415}, 54 (2002).

\bibitem{Ho09} A. Hosford, S. G. Ryan, A. E. Garcia-Perez, J. E. Norris and K. A. Olive, Astron. Astrophys. {\bf 493}, 601 (2009).

\bibitem{Bo10} R. N. Boyd, C. R. Brune, G. M. Fuller and C. J. Smith, Phys. Rev. {\bf D 82}, 105005 (2010).

\bibitem{Do93} B. D. Fields, S. Dodelson and M. S. Turner, Phys. Rev. {\bf D 47}, 4309 (1993).

\bibitem{Do94} S. Dodelson, G. Gyuk and M. S. Turner, Phys. Rev. {\bf D 49}, 5068 (1994).

\end{thebibliography}
\end{document}